\documentclass{article}
\usepackage{spconf,amsmath,graphicx}
\usepackage{amssymb}
\usepackage{mwe} 
\usepackage[colorlinks,linkcolor=blue]{hyperref}

\title{Dual Prompting for diverse count-level PET denoising}
 
\name{
\parbox{\linewidth}{\centering Xiaofeng Liu$^{1,2}$, Yongsong Huang$^1$, Thibault Marin$^{1,2}$, Samira Vafay Eslahi$^{3}$, Amal Tiss$^{3}$, Yanis Chemli$^{1}$, Keith A. Johnson$^{3}$, Georges El Fakhri$^{1,2}$, Jinsong Ouyang$^1$}
}

\address{$^1$Dept. of Radiology and Biomedical Imaging, Yale University, New Haven, CT, USA\\
$^2$Dept. of Biomedical Informatics and Data Science, Yale University, New Haven, CT, USA\\
$^3$Dept. of Radiology, Massachusetts General Hospital and Harvard Medical School, Boston, MA, USA\\
}

\begin{document}
%
\maketitle

\begin{abstract}
The to-be-denoised positron emission tomography (PET) volumes are inherent with diverse count levels, which imposes challenges for a unified model to tackle varied cases. In this work, we resort to the recently flourished prompt learning to achieve generalizable PET denoising with different count levels. Specifically, we propose dual prompts to guide the PET denoising in a divide-and-conquer manner, i.e., an explicitly count-level prompt to provide the specific prior information and an implicitly general denoising prompt to encode the essential PET denoising knowledge. Then, a novel prompt fusion module is developed to unify the heterogeneous prompts, followed by a prompt-feature interaction module to inject prompts into the features. The prompts are able to dynamically guide the noise-conditioned denoising process. Therefore, we are able to efficiently train a unified denoising model for various count levels, and deploy it to different cases with personalized prompts. We evaluated on 1940 low-count PET 3D volumes with uniformly randomly selected 13-22\% fractions of events from 97 $^{18}$F-MK6240 tau PET studies. It shows our dual prompting can largely improve the performance with informed count-level and outperform the count-conditional model.
\end{abstract}


\section{Introduction}

Positron Emission Tomography (PET) imaging plays a pivotal role in medical diagnostics, providing insights into metabolic and physiological processes. However, due to the constraints of radiation exposure, many clinical applications require PET imaging at lower doses, leading to a significant reduction in photon counts. This results in low-count PET images that suffer from high noise levels, degrading diagnostic accuracy~\cite{lu2019investigation}. Consequently, denoising low-count PET images has become a critical task~\cite{liu2019higher}. However, conventional deep learning methods often struggle with diverse count levels, and a generalizable model across different count levels is highly expected~\cite{xie2023unified,liu2024cross}.

\begin{figure}[t]
\begin{center}\vspace{+15pt}
\includegraphics[width=1\linewidth]{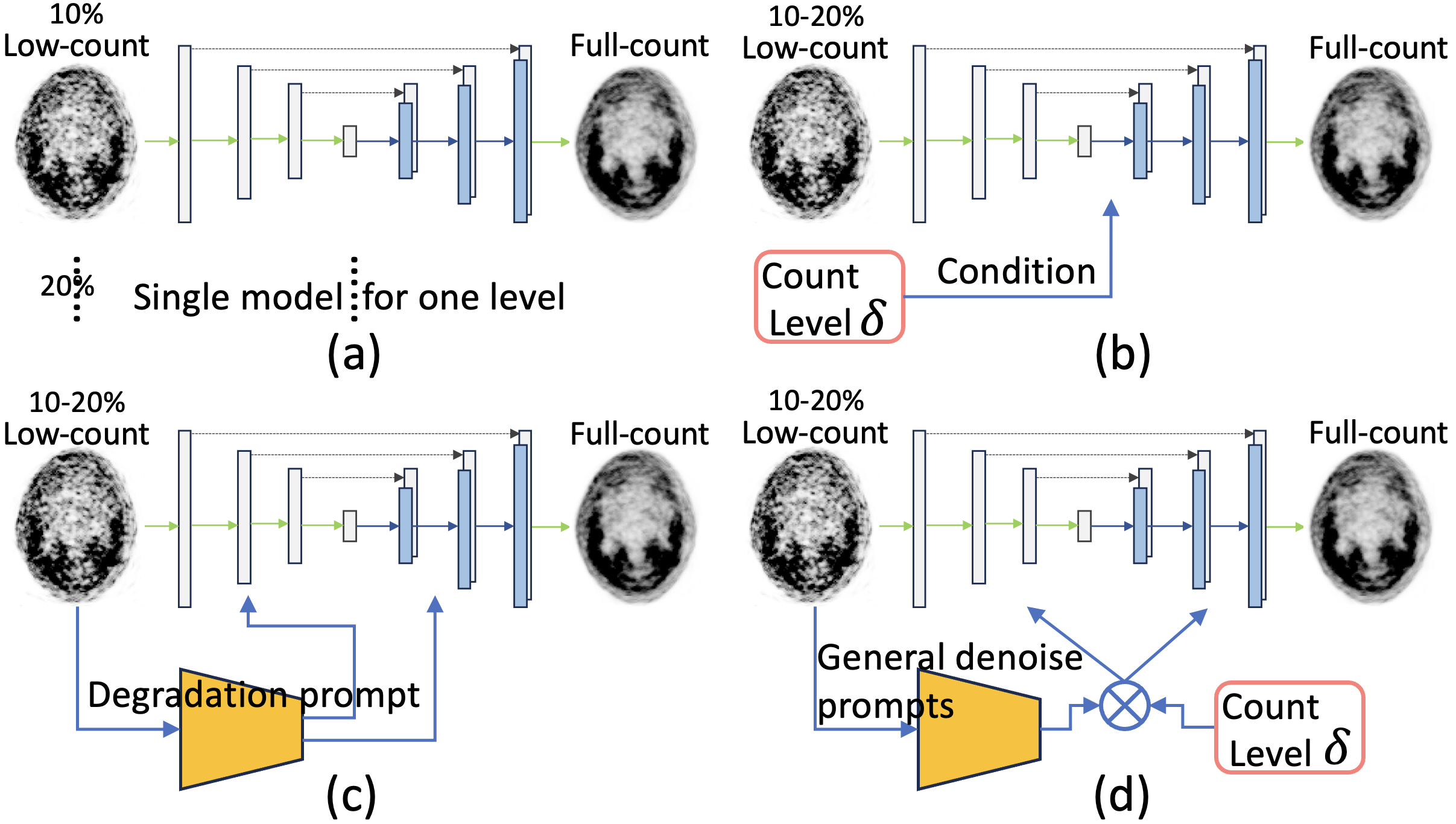}\vspace{-10pt}
\end{center}  
\caption{Illustration of applying (a) independent model~\cite{xie2023unified}, (b) conditional model~\cite{he2021interactive}, (c) blind prompting~\cite{potlapalli2024promptir}, and (d) our dual prompting for cross count-level PET denoising.} 
\label{fig1}\end{figure}

Recent work~\cite{xie2023unified} proposes to ensemble multiple models in discrete count levels. However, the count levels in real-world applications are continuous value~\cite{liu2024cross}. In addition, the independently trained model, as shown in Fig.1(a), also lacks interaction and flexibility for different count-level scans in their training. Instead,~\cite{liu2024cross} relies on adversarial training to achieve generalization on arbitrary unseen/unknown count levels, which is not able to utilize the count-level information in the testing stage. Though the count level can be quantified as prior knowledge to guide the PET denoising. An alternative solution can be adding the count value with a conditional module~\cite{he2021interactive} as shown in Fig.1(b). However, it is less effective to inject the information due to the degradation vanishing problem as investigated in~\cite{wang2023promptrestorer}.

\begin{figure*}[t!]
\begin{center} 
\includegraphics[width=1\linewidth]{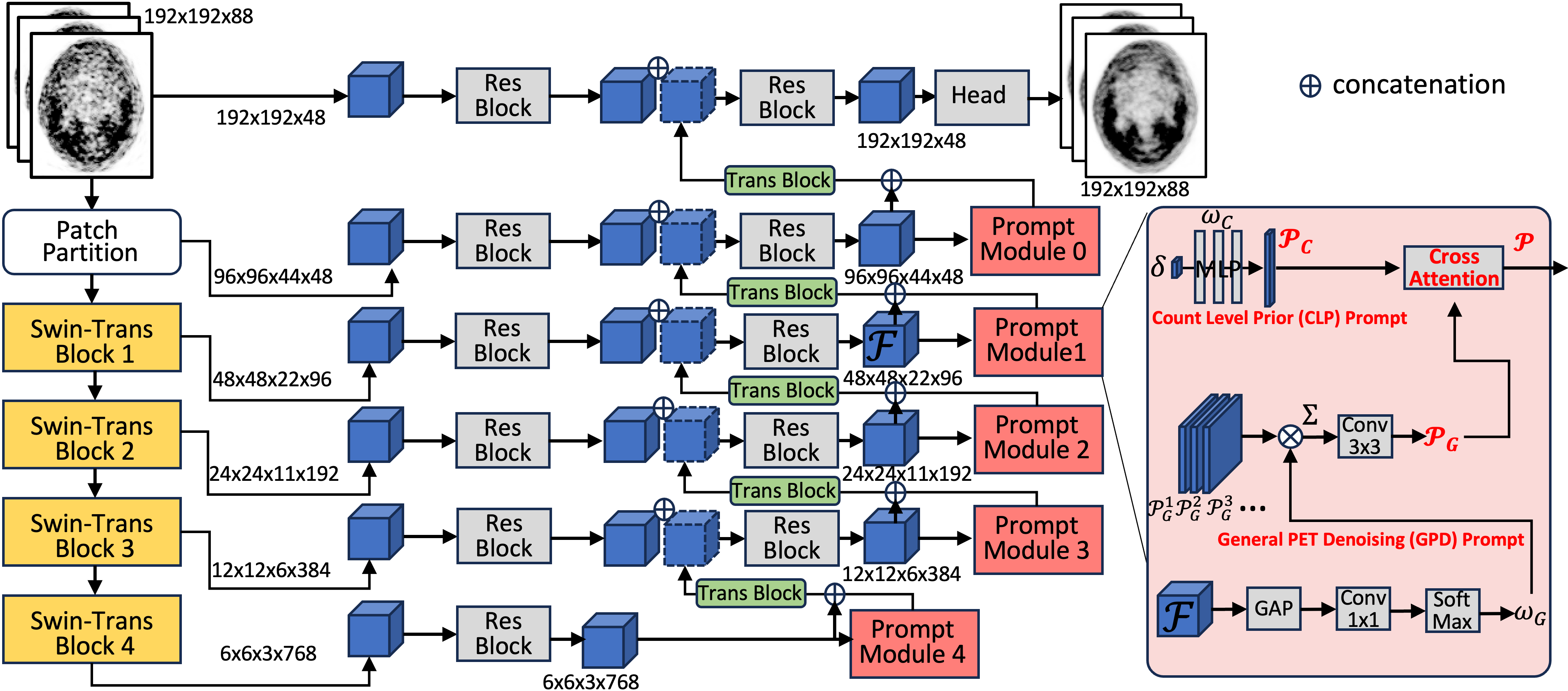}\vspace{-5pt}
\end{center}  
\caption{Illustration of our dual-prompt modules add-on to the 3D SWinUneter backbone. } 
\label{fig2}\end{figure*} 

Prompt learning was originally developed in large language models (LLMs) and text-to-image generation to equip the input data with a specific condition by encoding task-specific contextual information~\cite{lei2024prompt}.
The recent flourish prompt learning has been successfully adapted to achieve all-in-one image restoration with a unified model~\cite{ma2023prores,potlapalli2024promptir,wang2023promptrestorer}.  In image restoration, as illustrated in Fig.1(c), the learnable prompt is expected to encode crucial information of degradation type, which is then integrated with the image feature to dynamically guide the restoration of the image~\cite{potlapalli2024promptir}. Essentially, the prompt can guide the learned knowledge in a unified model to align with the specific input. Although prompt learning has recently shown promise in various natural image restoration, it still keeps a margin in medical image restoration, especially for 3D PET volumes.

In addition, compared to the linguistic prompt described with words, the learned prompt can usually include more sophisticated semantics reflecting the complicated image degradation~\cite{potlapalli2024promptir,wang2023promptrestorer}. However, relying on a module to learn all possible degradation information can be challenging for relatively limited training data in medical imaging. Notably, the count level in PET can be a known continuous value, which is different from the to-be-learned discrete degradation types in~\cite{potlapalli2024promptir,wang2023promptrestorer}.

Therefore, we propose to explicitly induce the prior knowledge of PET count-level as a specific prompt, while only implicitly encoding the general PET denoising knowledge with a learnable prompt. Then, the specific count-level prompt and general PET denoising prompt are integrated with a novel dual prompt fusion module. Our contributions can be summarized as follows:

\noindent$\bullet$ To our knowledge, this is the first attempt of prompt learning for medical image denoising. The special property of 3D PET with diverse count-level is further considered.

\noindent$\bullet$ A novel dual prompts scheme utilizes both the specific count-level prior prompt and learnable general PET denoising prompt to adaptively guide the denoising process.

\noindent$\bullet$A plug-and-play prompts integration module to seamlessly interact with the heterogeneous prompts and denoising features. It is potentially generalizable to U-shape models.

We demonstrate its effectiveness on 1940 low-count PET 3D volumes with uniformly randomly selected 13-22\% fractions of events from 97 $^{18}$F-MK6240 tau PET studies. It shows superior performance over the count-level conditional model~\cite{he2021interactive} and the recent learned prompt learning model~\cite{potlapalli2024promptir}.

\section{Methodology}

Given a low-count 3D PET volume \( \mathcal{X} \in \mathbb{R}^{H \times W \times D} \) (where \( H \), \( W \), and \( D \) denote height, width, and the number of slices respectively), the goal of the denoising network \( \mathcal{H}_{\theta} \) is to reconstruct a denoised image $\hat{\mathcal{Y}} = \mathcal{H}_{\theta}(X)$ to match its corresponding full-count PET volume \( \mathcal{Y} \in \mathbb{R}^{H \times W \times D} \). Our adaptive prompt guidance works as plug-and-play modules in the ship connections of the U-shape networks, each integrating the dual prompts as shown in Fig.~\ref{fig2}.

\subsection{Specific Count Level Prior (CLP) Prompt }

With the prior knowledge of count-level, we are able to configure the specific CLP prompt $\mathcal{P}_C\in \mathbb{R}^{M}$ to specifically guide the count-level relevant information. The scalar of court-level $\delta$ is a continuous value, which is mapped with multilayer perceptron (MLP) with three fully connected layers to a $M$-dimentional vector $\mathcal{P}_C=\omega_C(\delta)$ to better integrate with a high dimensional matrix of general PET denoising (GPD) prompt $\mathcal{P}_G$~\cite{karras2019style,liu2024treatment}.

\subsection{Learnable General PET Denoising (GPD) Prompt}

Recently, ProRes~\cite{ma2023prores} proposes adding a learnable prompt in the input image, while PromptIR~\cite{potlapalli2024promptir} further uses various prompts to modulate the feature map and demonstrates better performance in the natural 2D image restoration tasks. In this work, we follow~\cite{potlapalli2024promptir} to generate a learnable GPD prompt $\mathcal{P}_G$, which summarizes the other basic knowledge, e.g., the count-level irrelevant information in PET denosing. Compared with the word-wise prompt, the learnable prompts have remarkable flexibility and controllability. In low-level vision, describing the degradation with a few words is difficult due to the complexity of degradation. Concretely, the GPD generation module dynamically predicts attention-based weights from the input feature $\mathcal{F}\in \mathbb{R}^{h \times w \times c}$ and a set of random parameterized $\mathcal{P}_G^n\in \mathbb{R}^{N\times h \times w \times c}$:  
\begin{align}
    \omega_G &=\text{SoftMax}(\text{Conv}_{1\times1}(\text{GAP}(\mathcal{F}))),\in \mathbb{R}^{N},\\
    \mathcal{P}_G &=\text{Conv}_{3\times3}(\sum_{n=1}^N \omega_G \mathcal{P}_G^n), \in \mathbb{R}^{h \times w \times c},
\end{align} where GAP indicates global average pooling to generate a feature vector, which is followed by the downscaling convolution to encode a compact feature.

\subsection{Dual Prompts Integration}

Integrating two prompts with different shapes is a challenging task. We propose to adapt the cross-attention mechanism~\cite{chen2021crossvit,rombach2022high} for dual prompts fusion. Specifically, we have integrated prompt $\mathcal{P}$ following:
\begin{align}
   \mathcal{P} =  \text{SoftMax}(\textbf{Q}_G \textbf{K}_C^{\top}/\sqrt{d_k})\textbf{V}_C, \in \mathbb{R}^{h \times w \times c},
\end{align}
where the query $\textbf{Q}_G$ is derived from the $\mathcal{P}_G$ with layer normalization, and the key $\textbf{K}_C$ and value $\textbf{V}_C$ are derived from $\mathcal{P}_C$ processed by 1×1 convolutions and 3×3 depth-wise
convolutions, respectively. Finally, the resulting integrated PET denoising restoration prompt $\mathcal{P}$ serves as the prompt of features with both knowledge of count level and general PET noise degradation, which can be well understood by PET denoising models.

\subsection{Prompt Injection and Training Protocol}

The integrated prompt $\mathcal{P}$ is then channel-wise concatenated with the input feature $\mathcal{F}$, which is then processed by a transformer block~\cite{potlapalli2024promptir}. Notably, the transformer block involves two consecutive parts of a Multi-De-Conv head transposed attention (MHTA) for channel-wise correlation modeling, and a Gated-De-Conv feedforward layer (GFL) to compress the feature~\cite{zamir2022restormer}. Following~\cite{potlapalli2024promptir}, we empirically add the prompt in the skip connections of the UNet style model as detailed in Fig.~\ref{fig2}.

In training, we compute the corresponding denoising loss: 
\begin{align}
    \mathcal{L}_1=||\hat{\mathcal{Y}}-{\mathcal{Y}}||_1,
\end{align}
which measures the difference between the denoised image volume $\hat{\mathcal{Y}}$ and the full-count image volumes ${\mathcal{Y}}$. We update the network parameters in $\mathcal{H}_\theta$, $\omega_G$, $\omega_C$, and prompts integration module according to $\mathcal{L}_1$. In the testing stage, we input a low count 3D PET volume and its corresponding count level $\delta$ to generate its $\hat{\mathcal{Y}}$.

\begin{figure*}[t!]
\begin{center} 
\includegraphics[width=1\linewidth]{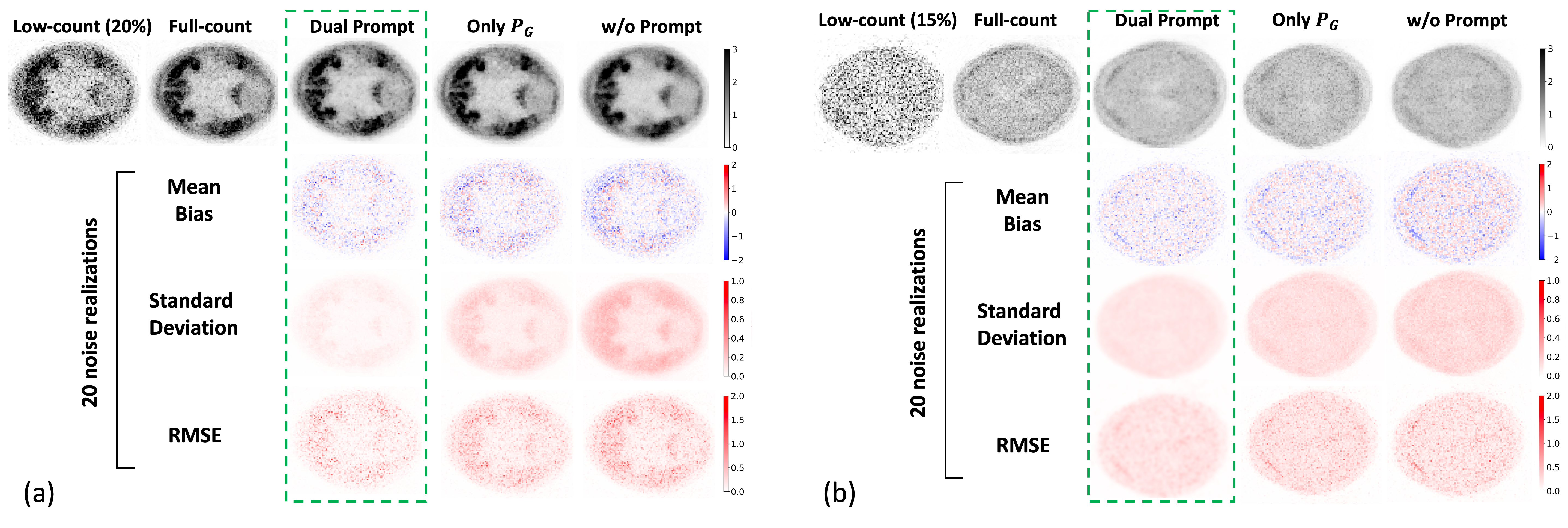}\vspace{-5pt}
\end{center}  
\caption{Comparison of the denoising with different methods for a PET slice in (a) 20\% count level and (b) 15\% count level, respectively. The voxel-wise mean bias, standard deviation, and RMSE are the lower the better.} 
\label{fig3}\end{figure*} 

\section{Experiments}

To evidence its effectiveness, we evaluate on PET studies with a novel tracer of $^{18}$F-MK-6240, which is designed to exhibit high binding affinity and selectivity for the tau protein~\cite{liu2024cross}. We utilized data from 60 participants, all of whom were scanned using the GE DMI PET-CT system. Specifically, we employed list-mode PET scans collected between 90 and 120 minutes post-injection. Since the PET data is in list mode, we reconstructed a subset of the data to generate a low-count image volume, while the full-count volume served as the reference. Additionally, each subject underwent an MRI scan using the MP-RAGE sequence on a Siemens 3T system.

The fraction of detected events is represented by $\delta$, with $\delta$ values ranging between 13\% and 22\%, selected uniformly. For each scan, 20 noise realizations were created by randomly subsampling events from the full-count list-mode data. The images were reconstructed using OSEM with five iterations and 16 subsets.

Standard Uptake Value Ratios (SUVRs) were calculated, using the cerebellum gray matter as a reference. Among the 60 subjects, some had multiple scans, resulting in 97 full-count scans in total. Consequently, the training, validation, and testing datasets were generated with 1400, 120, and 420 noise realizations, respectively, derived from 70, 6, and 21 PET studies using $^{18}$F-MK6240, ensuring subject-independent splits. Each reconstructed volume had dimensions of 192×192×88.

We implemented with Pytorch on a server with NVIDIA A5000 GPU. Without loss of generality, we use the 3D SwinUNeter~\cite{hatamizadeh2021swin} as our denoising model. Therefore, we have five prompt modules for its five skip connections. We set $N$ to 3 for a relatively small size. Notably, our contribution is orthogonal to more advanced backbones. We consistently trained 100 epochs with a batch size of 4 for all methods for a fair comparison.

For each test study, containing $K=$20 samples with $\delta\in[13,22]$, we computed the voxel-wise mean bias $m_i$, standard deviation $\sigma_i$, and root mean square error (RMSE) $rmse_i$, following the approach in \cite{ouyang2013bias}. Specifically, for voxel $i$, these metrics are given by:
\begin{align}
&m_i=\frac{1}{K}\sum_{k=1}^{K} (x^k_i-y_i),~~~\sigma_i=\sqrt{\frac{\sum_{k=1}^{K} (x^k_i-m_i)^2}{K-1}},\\ &rmse_i=\sqrt{m_i^2+\sigma_i^2},
\end{align}
where $x^k_i$ represents the denoised voxel value for the $k$-th sample, and $y_i$ is the corresponding full-count voxel value.

As shown in Fig.~\ref{fig3}, we compared two PET slices denoising performance of a single noise realization using 15\% or 20\% low-count, respectively. We can see that prompt learning is able to improve the performance of denoising in varied count levels. More appealingly, the additional specific CLP prompt can further boost the performance over the GPD prompt as~\cite{potlapalli2024promptir}.

In addition, we also assessed quantitative performance averaged over all 420 test samples. For each denoised sample, the mean absolute error (MAE), mean square error (MSE), 3D Structural Similarity Index Measure (SSIM), and Peak Signal-to-Noise Ratio (PSNR) were calculated with respect to the full-count reference. The average of these metrics across the 420 test samples was used for overall evaluation. The 3D SSIM metric~\cite{venkataramanan2021hitchhiker}, which accounts for structural integrity, contrast, and other perceptual elements, provides an assessment of perceived image quality, with higher values indicating greater structural similarity. We used regularization constants $K_1 = 0.01$, $K_2 = 0.03$, and applied a 3D Gaussian window with $k = 5$ and $\sigma = 1.5$ \footnote{https://github.com/jinh0park/pytorch-ssim-3D}. PSNR, on the other hand, quantifies noise and distortion levels, with higher values indicating less noise, though PSNR may not always align with perceived image quality in tasks like denoising or enhancement.

As shown in Table~\ref{tabel:1}, we compared the related methods and provided the ablation study of different prompts. It is worth noticing that we are investigating the continuous $\delta$, and not applicable to use the independent model-based fusion method~\cite{xie2023unified}. We also compared with the conditional model~\cite{he2021interactive} to inject the count-level, which is not benefited by the learned implicit GPT prompt. The inferior performance conditional model~\cite{he2021interactive} also aligns with the previous study in the natural image, which indicates its optimizable parameters result in gradually clearer features during the learning process, leading to the degradation vanishing which accordingly limits model performance. In addition, by adding the CLP prompt of $\delta$, our method outperforms the blind prompting method~\cite{potlapalli2024promptir} with only the learned GPD prompt.

\section{Conclusion}

In this work, we introduced a novel dual prompting framework for PET image denoising that leverages both priors of count-level and the learned general knowledge of PET denoising. The heterogeneous prompts are integrated with a cross-attention mechanism, and have been injected into the skip connection features in the U-shape denoising model. Through extensive experiments, we demonstrated that both CLP and GPD prompts can contribute to better performance in PET denoising with diverse count levels. It also presents a new paradigm for utilizing prior image degradation labels to guide the task-specific restoration over with conventional conditional models~\cite{he2021interactive}. Our proposed method can potentially be widely generalizable to other medical modalities with known or quantifiable degradation degrees.


\begin{table}[t]
\centering 
\caption{Numerical comparisons of different methods. $\uparrow$/$\downarrow$ the higher/lower, the better. $\dag$ This is the first time implementation on 3D medical data.} \vspace{+8pt}
\resizebox{1\linewidth}{!}{
\begin{tabular}{l|c|c|c|c}
\hline
Method &  MAE $\downarrow$  &  MSE $\downarrow$ &  PSNR $\uparrow$ &  3D SSIM $\uparrow$ \\\hline\hline

\text{Ours (Dual Prompts)} & \textbf{0.2205 }& \textbf{0.1031} & \textbf{19.9934} & \textbf{1.0941} \\\hline

Ours:only$\mathcal{P}_G$ (i.e.,\cite{potlapalli2024promptir}$^{\dag}$)& 0.2253 & 0.1059 & 19.4504 & 1.0889 \\ \hline

Ours:only$\mathcal{P}_C$ & 0.2376 & 0.1108 & 19.1420 & 1.0846 \\ \hline\hline

Condition-$\delta$~\cite{he2021interactive}$^{\dag}$ & 0.2401 & 0.1199 & 19.0949 & 1.0830 \\ \hline

\text{SWinUneter (direct train)} & 0.5726 & 1.3381 & 13.9907 & 0.8977 \\ \hline

\text{Unprocessed low count} & 0.7132 & 1.3558 & 11.8158 & 0.7374 \\\hline

\end{tabular}} 
\label{tabel:1}  
\end{table}

\newpage
  
\section{COMPLIANCE WITH ETHICAL STANDARDS} 
This research study was approved by the Institutional Review Board of Massachusetts General Hospital at Harvard Medical School.

\section{ACKNOWLEDGMENTS} 
This work is supported in part by NIH grants P41EB022544, R21EB034911, R01CA275188, R01AG076153, T32EB013180, and P01AG036694.

\bibliographystyle{IEEEbib}
\bibliography{refs}

\end{document}